\documentclass[pre,showpacs,amssymb,preprint]{revtex4}
\usepackage{amsmath,graphicx}
\usepackage{epsfig}

\begin{document}

\title{Anomalous diffusion of a tethered membrane: A Monte Carlo
investigation} 

\author{Hristina Popova and Andrey Milchev}

\affiliation{
Institute of Physical Chemistry, Bulgarian Academy of Sciences, $1113$
Sofia, Bulgaria}

\begin{abstract}

Using a continuum bead-spring Monte Carlo model, we study the anomalous
diffusion dynamics of a self-avoiding tethered membrane by means of extensive
computer simulations. We focus on the subdiffusive stochastic motion of the
membrane's central node in the regime of flat membranes at temperatures above
the membrane folding transition.
While at times, larger than the characteristic membrane relaxation time
$\tau_R$, the mean-square displacement of the center of mass of the sheet,
$\langle R_c^2\rangle$, as well as that of its central node, $\langle
R_n^2\rangle$, show the normal Rouse diffusive behavior with a diffusion
coefficient $D_N$ scaling as $D_N \propto N^{-1}$ with respect to the number of
segments $N$ in the membrane, for short times $t\le \tau_R$ we observe a {\em
multiscale dynamics} of the central node, $\langle R_n^2\rangle \propto
t^\alpha$, where the anomalous diffusion exponent $\alpha$ changes from $\alpha
\approx 0.86$ to $\alpha \approx 0.27$, and then to $\alpha \approx 0.5$, before
diffusion turns eventually to normal.
By means of simple scaling arguments we show that our main result, $\alpha
\approx 0.27$, can be related to particular mechanisms of membrane dynamics
which involve different groups of segments in the membrane sheet. A comparative
study involving also linear polymers demonstrates that the diffusion coefficient
of self-avoiding tethered membranes, containing $N$ segments, is three times
smaller than that of linear polymer chains with the same number of segments.
\end{abstract}
\maketitle

\section{Introduction}

There has been considerable interest recently in understanding the
statistical properties of polymerized (or tethered) membranes \cite{Gompper}.
This interest in large part is due to the membrane behavior which is much richer
than that of polymers, their one-dimensional analog. In addition, this interest
is justified by a variety of real systems like red-blood-cell
cytoskeletons \cite{Schmidt}, graphite oxide sheets \cite{Hwa,Spector} or
dispersed silicate (clay) platelets \cite{Sinsawat,Vaia} which can be modeled by
networks of fixed connectivity, generally referred to as polymerized membranes.
Along with the experimental studies, self-avoiding polymerized membranes have
also attracted remarkable interest from the point of view of basic research in
recent years. Their static properties have been studied analytically and
numerically
\cite{KKN,Doussal,Liu,YKKK,Radzihovsky,Plishke,Boal,FAbrah,DWH,Mori,Drovetsky,
Kownacki,Koibuchi}. Much of these studies have been spent in the pursuit of the
so called ``crumpling transition'' between a low-temperature flat phase and a
high-temperature crumpled phase until it was realized
\cite{FAbrah,Grest,GompperKroll} that self-avoiding membranes are always
flat (with an {\em infinite} persistence length), i.e., their radius of gyration
$R_g$ scales with linear size $L$ as $R_g \propto L^\nu$ where the Flory
exponent $\nu \approx 1$. The flat phase arises even without explicit bending
rigidity because the resistance to in-plane shear deformations leads to
anomalous stiffening of the surface in the presence of thermal fluctuations. 

In contrast to static properties, the membrane dynamics is less well-understood.
Earlier analytical and numeric studies \cite{KKN,Boal,Frey,Wiese} have revealed
that the self-avoiding restrictions considerably modify the relaxation times of
the tethered surface. Thus the typical relaxation time $\tau_R$ of a tethered
membrane in the case of Rouse dynamics when hydrodynamic interactions are
neglected has been predicted by simple scaling arguments \cite{KKN} to vary as
$\tau_R \propto L^{2+2\nu} \propto R_g^{2+2/\nu}$. If, as in polymer physics,
one introduces a {\em dynamic exponent} $z$, describing the relaxation process
as $\tau_R \propto R_g^z$, then one gets $z = 2+2/\nu$ (for linear polymers one
has $z = 2+1/\nu$). Usually, $\tau_R$ is considered to be the time needed for
the membrane to diffuse its radius of gyration. For tethered membranes, highly
permeable to the solvent as in isolated spectrin networks, one is in the Rouse
regime \cite{Frey} and the diffusion coefficient $D_N$ scales with membrane size
$L$ as $D_N\propto L^{-2}$. Thus the time it takes for such a flat membrane to
move a distance $R_g$ is proportional to $L^4$. In contrast, for impermeable
membranes (like, e.g., erythrocytes) where solvent backflow (i.e., a long-ranged
hydrodynamic interaction) is important, one has in $d$-dimensions in the case of
Zimm dynamics $\tau_H\propto R_g^d\propto L^{d\nu}$ (i.e. $z=d$) and
$D_N\propto L^{-1}$. Thus one may view permeability as constituting two
different dynamic universality classes of tethered membranes \cite{Frey} whereby
these classes (Rouse dynamics - highly permeable membranes, or Zimm dynamics -
impermeable membranes) are observed for a wave vector independent (or,
dependent) friction coefficient. Recently, a series of simulation studies by
Pandey et al. \cite{Pandey,Pandey_old} has revealed a multiscale stochastic
dynamics of tethered membranes at times before normal diffusive behavior is
reached. The displacement motion of the central node, $R_n$, of a
four-coordinated coarse-grained model membrane has been observed to undergo a
subdiffusive mean-square displacement (MSQD) $\langle R_n^2\rangle \propto
t^\alpha$ with the exponent $\alpha$ attaining different values in the short and
intermediate time regimes before turning to normal diffusion for $t \ge \tau_R$
with $\alpha = 1$. 

In the present work we employ an efficient off-lattice Monte Carlo algorithm,
focusing on the subdiffusive dynamics of self-avoiding tethered membranes and
comparing some of the salient dynamic features to those of linear polymers. Our
observations, based on extensive computer simulations, largely confirm those of
previous investigators \cite{Pandey,Pandey_old}. As a step forward, however, we
suggest a scaling theory which explains our findings for the anomalous membrane
dynamics, relating the observed values of $\alpha$ to the specific stochastic
motion  of particular groups of sheet segments. 

After briefly introducing our model in Section II, this is considered in Section
III where we focus on the main results of our investigation and their
interpretation. We close this paper with a brief summary and discussion in
Section IV.

\section{Model system and simulation procedure}

\begin{figure}[htb]
\includegraphics[scale=0.3, angle=0]{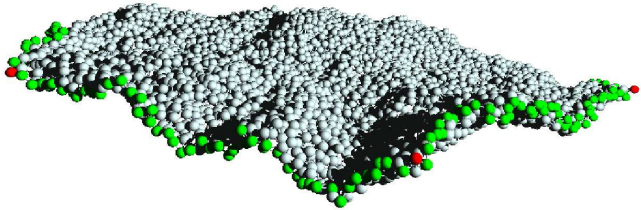}
\caption{(Color online) A snapshot of a tethered membrane at $T = 1.0$ with
linear size (the edge length of a regular hexagonal sheet) $L=50$ which contains
$N=7351$ monomers. Periphery segments at the rim of the membrane are shaded grey
(green) while the six monomers at the vertices (corners) of the sheet are dark
grey (red).}
\label{snap}
\end{figure}

We study a coarse-grained model of \emph{self-avoiding tethered membranes},
embedded in three-dimensional space. The membranes have a hexagonal lattice
structure where each monomer interacts with six nearest-neighbors -
Fig.~\ref{snap}. There are altogether $N=(3L^2-3L+1)$ monomers in
such a membrane where by $L$ we denote the number of monomers on the edge of the
network (i.e. $L$ is the linear size of the membrane). In this model, spherical
particles of diameter $\sigma$ are connected in a fixed geometry by flexible
strings of length $l$. To prevent self-intersection of the membrane, the maximum
length of the strings between the centers of the spheres must be
$l\leq\sqrt{3}\sigma$, then the membrane is self-avoiding in that it cannot
intersect itself.

The bonded nearest neighbor monomers on the membrane interact with each other
through the finitely extensible nonlinear elastic (FENE) potential
\cite{Bird} where a bond $l$ has a maximum length $l_{\text{max}}$ and a minimum
length $l_{\text{min}}$,
\begin{equation}
U_{\text{FENE}}(l) = -K(l_{\text{max}}-l_0)^2 \ln \left[
1-\left(\frac{l-l_0}{l_{\text{max}}-l_0}\right) ^2 \right] .
\label{eq_U_FENE}
\end{equation}
The minimum of this potential occurs for $l=l_0$ , $U_{\text{FENE}}(l_0)=0$,
near $l_0$ it is harmonic, with $K$ being a spring constant, and the potential
diverges to infinity both when $l\rightarrow l_{\text{max}}$ and when
$l\rightarrow l_{\text{min}}$. Choosing our length unit $l_{\text{max}}=1.0$, we
choose the other parameters as $ l_{\text{min}}=0.2 ,~
l_0=(l_{\text{min}}+l_{\text{max}})/2=0.6 ,~ K/k_BT=5 $, where $T$
denotes the absolute temperature, and $k_B$ is the Boltzmann constant.

Self-avoidance is observed by the interaction between particles which are not
nearest neighbors on the network. The nonbonded interaction between monomers is
described by a Morse potential where $r$ is the distance between the monomers,
\begin{equation}
U_{\text{Morse}}(r)/\epsilon_M=\exp[-2\alpha(r-r_{\text{min}})]-2\exp[
-\alpha(r-r_{\text{min}} ) ]
\label{eq_U_MORSE}
\end{equation}
with parameters  $\epsilon_M/k_BT=1,~ \alpha=24$. The minimum of this potential
occurs for $ r=r_{\text{min}},~ U_{\text{Morse}}(r_{\text{min}})/\epsilon_M=-1
$. For $\alpha=24,~ U_{\text{Morse}}(r)$ essentially is zero for $r\ge
1.25~r_{\text{min}}$. Choosing then units of length such that
$r_{\text{min}}=0.8$, we hence can take $U_{\text{Morse}}(r\geq1)=0$. The
repulsive part of this potential guarantees self-avoidance of the membrane.

We have used the standard Monte Carlo procedure to investigate the thermodynamic
properties of self-avoiding tethered membranes. The total energy (Hamiltonian)
is the sum of Eqs.~(\ref{eq_U_FENE}) and (\ref{eq_U_MORSE}). In each Monte
Carlo update, a monomer is chosen at random and one attempts to displace it
randomly by displacements chosen uniformly from the intervals ${ -0.25 \leq
\Delta x,\; \Delta y,\; \Delta z \leq +0.25 }$. The attempted move is accepted
or rejected according to the conventional Metropolis criterion by
comparing the transition probability $W=\exp(-\Delta U/k_BT)$ (where $\Delta U$
is the energy difference between the configurations after and before the trial
move) with a random number uniformly distributed between zero and unity. If
$W$ exceeds this random number, the attempted move is accepted, otherwise
it is rejected. Time is measured in Monte Carlo steps (MCS) per monomer whereby
a single MCS is elapsed after $N$ monomers are picked at random and given the
chance to perform a trial move. Since our potentials are constructed such that
the membrane cannot intersect itself in the course of random displacements of
monomers, one does not need to check separately for entanglement restrictions
during the simulation. Thus the algorithm is reasonably fast. Nevertheless, the
simulation takes quite a long time for large self-avoiding membranes to 
equilibrate and then move a substantial distance in space. This and the
necessity to attain very good statistical accuracy have limited our
investigations to sizes $L\le 50$. Eventually, we would like to note that the
interactions used in the present off-lattice model, albeit somewhat more refined
and complicated than the simple potential used in earlier simulations on a cubic
lattice~\cite{Pandey_old,Pandey}, do not change the physics of the problem and
lead qualitatively to the same results.

\section{Results}

\begin{figure}[htb]
\includegraphics[scale=0.4, angle=270]{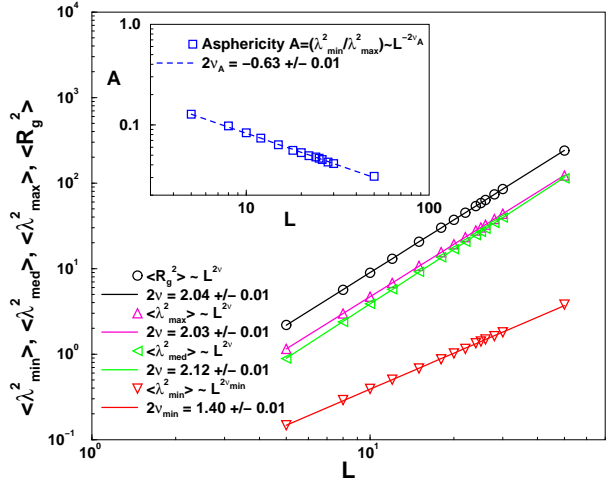}
\caption{(Color online) Log-log plot of the mean-squared radius of gyration,
$R_g^2$, and the three eigenvalues,
$\lambda_{\text{max}}^2,\;\lambda_{\text{med}}^2,\;\lambda_{\text{min}}^2$
against the linear size $L$ of flat tethered membranes ($5\le L\le 50$) at $T =
1.0$. The respective exponents are given in the legend. In the inset we show the
variation of membrane asphericity, $A=\lambda_{\text{min}}^2 /
\lambda_{\text{max}}^2$ which vanishes with $L$ as $A\propto L^{-0.63}$.}
\label{gyration}
\end{figure}

Before we focus on the subdiffusive dynamics of our membranes, we show in Fig.
\ref{gyration} the scaling behavior of the gyration radius, 
\begin{equation}
\langle R_g^2 \rangle= \frac{1}{N} \sum \limits_{i=1}^N \left\langle
(r_i-r_{cm})^2 \right\rangle ,
\label{eq_Rg}
\end{equation}
where $r_i$ is the position of the $i$th monomer of the membrane while $ r_{cm}
= \frac{1}{N} \sum \limits_{i=1}^N r_i $ is its center of mass location.
We also sample the eigenvalues $\lambda_{\text{max}}^2,\;
\lambda_{\text{med}}^2,\; \lambda_{\text{min}}^2$ of the inertial tensor,
\begin{equation}
I_{\alpha\beta}=\frac{1}{N} \sum \limits_{i=1}^N (r^{\alpha}_i -
r^{\alpha}_{cm})(r^{\beta}_i- r^{\beta}_{cm}) ,
\label{eq_InertiaTensor}
\end{equation}
where $\alpha,\beta \in \lbrace x,y,z \rbrace$, the sum is taken over all
particles of a given configuration, and $r^{\alpha}_{cm}$ is the $\alpha$
component of the center of mass radius vector for a given configuration.  The
three eigenvalues are ordered according to magnitude $ \lambda_{\text{max}}^2
\geq \lambda_{\text{med}}^2 \geq \lambda_{\text{min}}^2 $. The directions of the
principal axes are given by the three eigenvectors corresponding to the three
eigenvalues. For a planar membrane, the eigenvector associated with
$\lambda_{\text{min}}^2$ is perpendicular to the plane of the membrane while the
eigenvectors associated with $\lambda_{\text{max}}^2$ and
$\lambda_{\text{med}}^2$ lie in the plane of the membrane. It is evident from
Fig.~\ref{gyration} that our membranes are indeed flat with scaling exponents
$\nu = 1.02\pm 0.01$ for $R_g$, $\nu = 1.02 \pm 0.01$ for
$\lambda_{\text{max}}^2$, and $\nu = 1.06\pm 0.01$ for $\lambda_{\text{med}}^2$.
The asphericity ratio $A=(\lambda_{\text{min}}^2/\lambda_{\text{max}}^2)\propto
L^{-2\nu_A}$ tends to zero with an exponent $\nu_A = 0.32\pm 0.01$, indicating
that these membranes are indeed asymptotically flat. Note that these data have
been obtained at $T=1.0$ well above the temperature of the first folding
transition \cite{JCP}, $T_{c_1}=0.89\pm 0.01$. In Fig.~\ref{gyration} and in
the following figures the error bars do not exceed the size of the symbols.

Turning now to membrane dynamics in the Rouse regime, one may assume that each
segment of the membrane moves under the influence of surface forces (surface
stretching due to near-neighbors and excluded volume forces due to distant
neighbors), and a random force representing thermal noise. As far as the
contribution of inertial terms to membrane motion can be neglected for
sufficiently long times, one may assume that the relevant dynamics is purely
diffusive. With $\rho$ being the rate of position changes of monomers per unit
time and $z=2/\nu+2$, the dynamic exponent, one may write the relaxation time
of the membrane $\tau_R$ as
\begin{eqnarray}
\tau_R = \rho^{-1}R_g^z = \rho^{-1}N^{z\nu/2} 
\end{eqnarray}
If monomeric orientations add up randomly and one neglects correlations, the
MSQD of the membrane center of mass is
\begin{eqnarray}\label{g3}
g_3(t) = \langle [\vec{r}_{cm}(t) - \vec{r}_{cm}(0)]^2\rangle = \rho \langle
\left(\frac{l}{N}\right)^2 \rangle Nt = \rho \frac{\langle \text{\em
l}^2\rangle}{N}t, 
\end{eqnarray}
because each monomeric motion moves the center of mass by a random displacement
of the order {\em l}$/N$, {\em l} being the bond length. There are $\rho N$ such
random motions per unit time. Invoking the Einstein relation $g_3(t) = 2d D_N t$
(where $d$ is the spatial dimensionality), one thus concludes 
\begin{eqnarray}\label {D_N}
D_N \propto \rho \langle \text{\em l}^2\rangle/N.
\end{eqnarray}
\begin{figure}[htb]
\includegraphics[scale=0.4, angle=270]{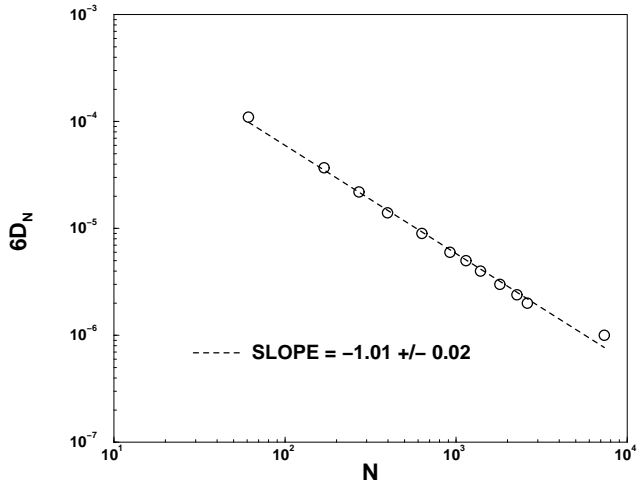}
\caption{(Color online) Variation of the diffusion coefficient $D_N$ of a
membrane with the number of monomers $N$ for membrane sizes $5 \le L \le 50$ at
$T = 1.0$. The measured slope is $-1.01 \pm 0.02$.}
\label{D}
\end{figure}
From Fig. \ref{D} it is evident that this prediction, Eq.~(\ref{D_N}), is
indeed nicely confirmed by the simulation. The relaxation time $\tau_R$ and the
scaling law $z\nu = 2\nu + 2$ is then understood by the condition that the
membrane is relaxed when its center of mass has diffused over its own size
$R_g$, i.e., 
\begin{eqnarray}
g_3(\tau_R) \propto \rho \frac{\langle \text{\em l}^2\rangle}{N} \tau_R \propto
\langle \text{\em l}^2\rangle N^{z\nu/2-1} \propto R_g^2 \propto \langle
\text{\em l}^2\rangle N^\nu
\end{eqnarray}
whence $z\nu = 2 +2\nu$ follows.

These arguments can be carried over for the MSQD of the membrane central node
as well. We define 
\begin{eqnarray}
g_1(t) = \langle [\vec{r}_n(t) - \vec{r}_n(0)]^2 \rangle
\propto \langle \text{\em l}^2\rangle (\rho t)^\alpha
\end{eqnarray}
at time $t < \tau_R$ anticipating that the central node exhibits anomalous
diffusion with an exponent $\alpha < 1$. For short times $(\rho t) \le 1$, of
course, a nearly free diffusion of the central node takes place, and thus
$g_1(t)$ for $\rho t \approx 1$ should be of the order of $\propto \langle
\text{\em l}^2\rangle$. Requiring now that $g_1(\tau_R) \approx R_g^2$, one gets
a scaling relation for $\alpha$,
\begin{eqnarray}\label{beta}
g_1(\tau_R) \propto  \langle \text{\em l}^2\rangle (\rho \tau_R)^\alpha \propto
\langle \text{\em l}^2\rangle N^{\alpha z\nu/2 } \propto \langle \text{\em
l}^2\rangle N^\nu.
\end{eqnarray}
Thus for flat membranes with $\nu = 1$, one has $\alpha ^{-1} = 1 + \nu^{-1} =
2$, and one would then expect to see a time interval $t < \tau_R$ where $g_1(t)
\propto t^{1/2}$. Additional information for the subdiffusive dynamics of the
membrane may be obtained if one defines in analogy with the case of linear
polymers \cite{AMKBJB} the MSQD of a central node measured in the center of mass
coordinate system of the membrane,
\begin{eqnarray}
 g_2(t) = \langle [\vec{r}_n(t) -\vec{r}_{cm}(t) -
\vec{r}_n(0) + \vec{r}_{cm}(0)]^2\rangle ,
\end{eqnarray}
and also for the averaged MSQD of the six monomers at the vertices of
the hexagonal sheet in the laboratory system of coordinates,
\begin{eqnarray}
g_4(t) = \langle \frac{1}{6} \sum_{i=1}^6 [\vec{r}_i(t) - \vec{r}_i(0)]^2 
\rangle .
\end{eqnarray}
In the center of mass coordinate system of the membrane,
\begin{eqnarray}
g_5(t) = \langle \frac{1}{6} \sum_{i=1}^6 [\vec{r}_i(t) -\vec{r}_{cm}(t) -
\vec{r}_i(0) + \vec{r}_{cm}(0)]^2 \rangle .
\end{eqnarray}
Evidently, for $t < \tau_R$ one should observe $g_2(t) \approx g_1(t)$ and
$g_5(t) \approx g_4(t)$ whereas $g_2(t) \propto R_g^2$ for $t \gg \tau_R$ since
the central monomer cannot travel farther from the center of mass than the
membrane size, of course.

\begin{figure}[htb]
\includegraphics[scale=0.4, angle=270]{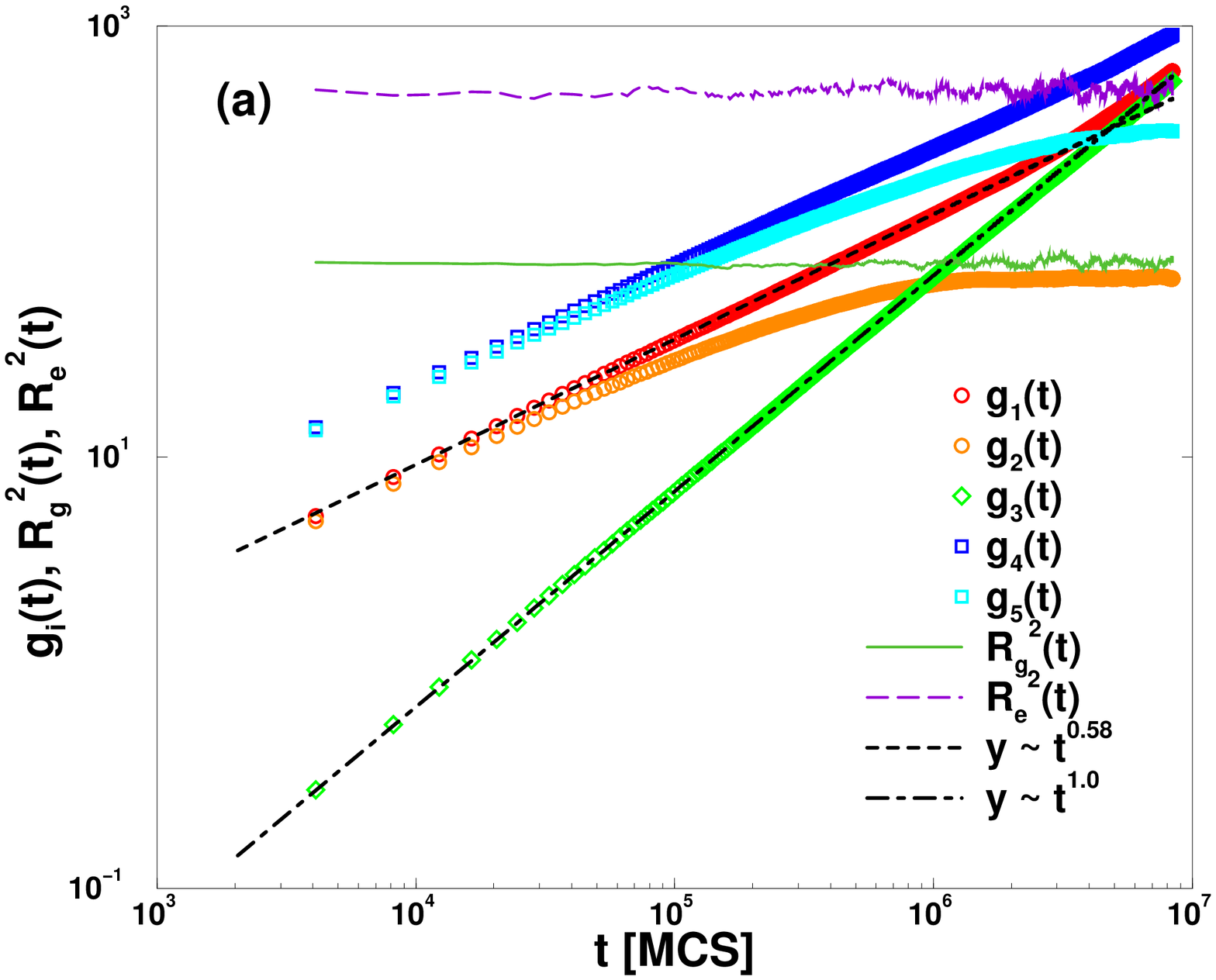}
\includegraphics[scale=0.4, angle=270]{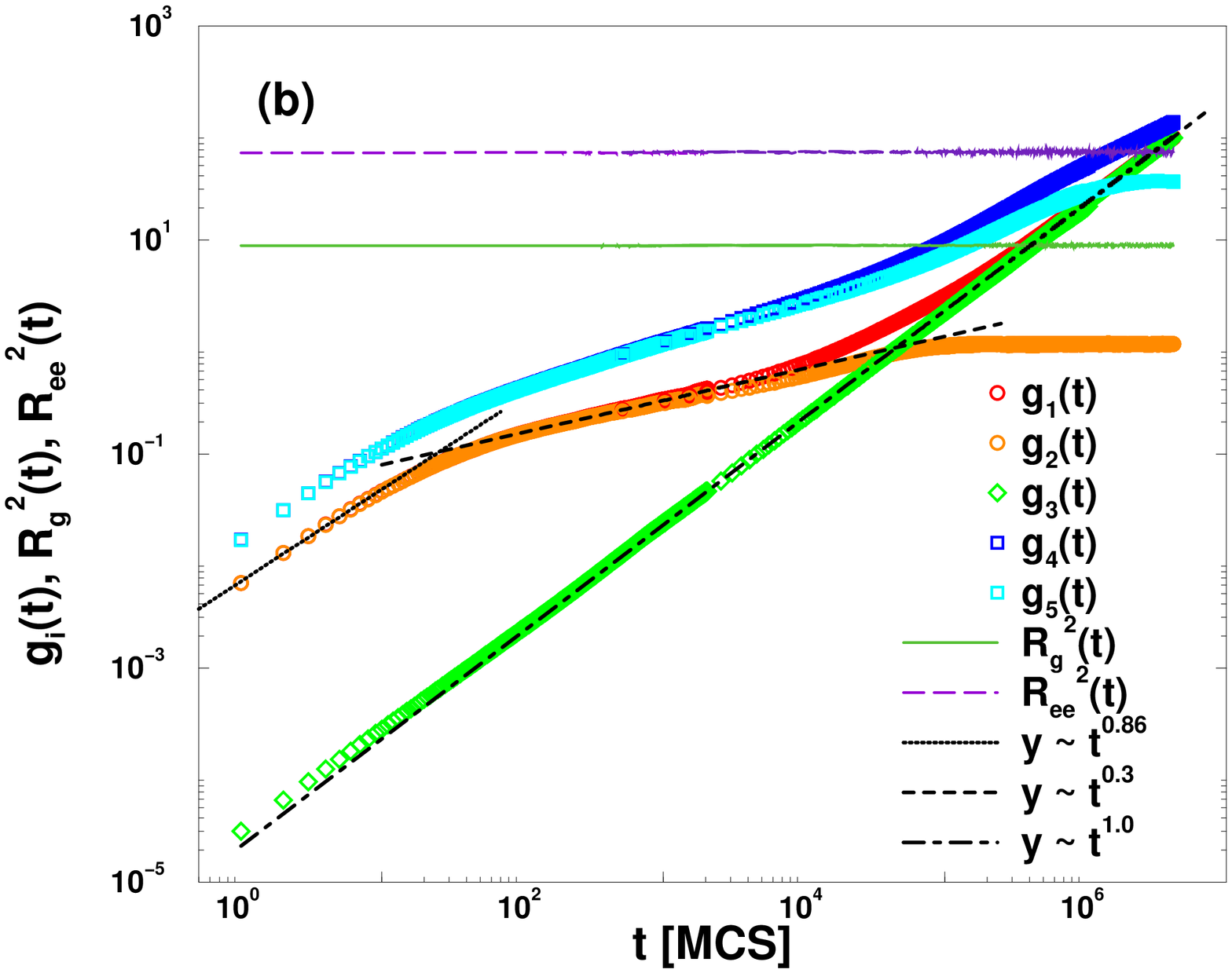}
\caption{(Color online) Log-log plot of mean square displacements $g_1(t),\;
g_2(t),\; g_3(t),\; g_4(t),\; g_5(t)$  at $T=1.0$ plotted vs time $t$ (measured
in Monte Carlo steps) for a linear polymer of length $N=256$ (a), and for a
membrane of size $L=10$ which contains $N=271$ monomers (b).
Dashed lines indicate the scaling behavior of the central segment MSQD,
$g_1(t)\propto t^{\alpha}$, and of the center of mass, $g_3(t)\propto t$,
with elapsed time while horizontal lines denote the time averages of the
radius of gyration $R_g^2$ and of the ``end-to-end'' distance of the chain,
$R_e^2$ (a), and of the membrane, $R_{ee}^2$ (b), the latter being measured
as the distance between the opposite vertices of the hexagonal sheet. Evidently,
cf. (b), by defining $\tau_R$ from $g_3(\tau_R)=R_g^2$ as the mean relaxation
time of the membrane one can verify that $g_1(\tau_R) \approx g_3(\tau_R)$. }
\label{MSQD}
\end{figure}

A general impression about the time variation of the various MSQD $g_i(t)$ and
the similarity in the stochastic dynamics of linear polymers and tethered
membranes may be gained from Fig. \ref{MSQD}. 
It is seen that the course of the functions $g_i$ with time is qualitatively
very similar for both linear polymers and tethered membranes. In both cases, cf.
Figs. \ref{MSQD}(a) and \ref{MSQD}(b), one finds that the center of mass
performs normal diffusion
with $g_3(t) = 6D_Nt$. One can, therefore, compare the relative diffusivity of
polymers and membranes, containing the {\em same} number of monomers $N$, say a
chain with $N=256$ and a membrane with $L=10$, i.e., $N=271$, with identical
forces acting between the repeating units. Our analysis shows that in a good
solvent, $T = 1.0$, one obtains $6D_N(\text{polymer})=6.6\times 10^{-5}$
and $6D_N(\text{membrane}) = 2.2\times 10^{-5}$, i.e., a linear self-avoiding
chain of $N$ segments moves {\em three times faster} than a self-avoiding
flat sheet in the case of Rouse dynamics. We find this result rather remarkable
since this decrease in mobility is solely and entirely due to the higher
topological dimensionality of the membrane.

A marked difference between  chains and membranes, however, is revealed if one
looks at the subdiffusive behavior of the central monomer in both cases. For
times shorter than the typical relaxation time, $t < \tau_R$, the central node
of the polymer chain is observed to diffuse like $g_1(t) \propto t^{0.58}$ (i.e.
very close to the expected $t^{0.54}$ power law) while for the membrane one
finds a much smaller power $g_1(t) \propto t^{0.3}$, seen also by Pandey et
al. \cite{Pandey_old}. It might be argued that this small exponent $\alpha
\approx 0.3$, describing the subdiffusive behavior of a tethered membrane,
reflects a membrane-specific dynamic mechanism which shows up at $t < \tau_R$.
In the following we suggest a possible interpretation and a simple scaling
derivation for the observed value of this novel exponent $\alpha$.
\begin{figure}[htb]
\includegraphics[scale=0.4, angle=270]{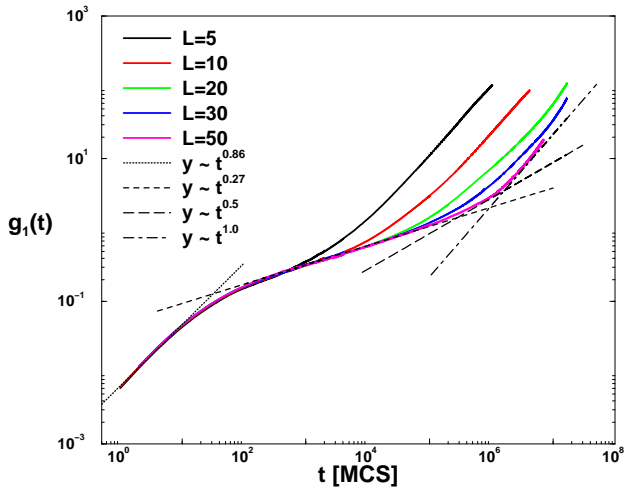}
\caption{(Color online) Log-log plot of the central node MSQD $g_1$ vs time $t$
at $T=1.0$ for a membranes with linear size $5\le L\le 50$. All data are
averaged over $100$ simulation runs. Dashed lines denote power law
variation with different exponents (see legend)  corresponding to the various
subdiffusive regimes.}
\label{g1}
\end{figure}

We first look more closely at the main data of our study showing the MSQD of the
central monomer of a tethered membrane in Fig.~\ref{g1}. The different regimes
of subdiffusive motion of the central node are indicated by power laws with
exponents, specified in the legend of Fig.~\ref{g1}. It is seen that for very
short time, $0 < t \le 1$, each segment indeed performs displacements which
are not constrained by the topological connectivity of the network so with
$g_1(t) \propto t^{0.86}$ one observes an extremely shortlived nearly normal
diffusion. At late times, $t \ge \tau_R$, the normal diffusive motion sets on
eventually, and $g_1(t) \approx g_3(t) \propto 6D_Nt$. We should like to point
out that at late times the averaging of the correlation functions $g_1$ for the
two largest system sizes, $L=30 \div 50$, is not perfect due to a progressively
deteriorating statistics, however, it is beyond doubt that their ultimate slope
corresponding to normal diffusion should be unity. In the intermediate time
interval our data yields a subdiffusive motion of the central monomer with
$g_1(t)\propto t^\alpha$ where $\alpha = 0.27 \pm 0.01$. Due to strong
finite-size effects this value of the $\alpha$ can be unambiguously established
for sufficiently large, $20 \le L \le 50$ membranes only. We note that very
close values for $\alpha \approx 0.25 \div 0.32$ have been observed recently in
the computer experiments of Pandey et al. \cite{Pandey_old} for the case of
tethered membranes in the good solvent regime at temperatures $2.0 \le T\le 10$.
Such behavior cannot be explained by means of the exponent $\alpha = 0.5$ which
follows from the estimate, Eq.~(\ref{beta}).

We believe that a possible explanation of this sluggishness of flat membranes at
early times $1 < t \le \tau_1 < \tau_R$ may be found if one assumes that in
this interval only the most loosely bound monomers (those at the membrane
periphery, or rather, those at the six vertices of the hexagonal sheet) actually
contribute to a displacement of the membrane center of mass while all monomers
with six-fold coordination in the bulk of the membrane are virtually blocked by
their neighbors and for this short time hardly move. As far as the membrane
retains its flat shape and does not fold, the maximal displacement of these
loosely bound monomers cannot exceed the effective thickness
$\lambda_{\text{min}}$. Indeed, a comparison of Fig.~\ref{gyration} and
Fig.~\ref{g1} shows that the MSQD, performed by a membrane of linear size $L$
during the time $\tau_1$ (the latter is given by the intersection point of the
tangent $y \propto t^\alpha$ to $g_1$, and $g_3 \propto t$) amounts to
$g_3(\tau_1) \approx \lambda_{\text{min}}^2$. One can, therefore, estimate the
characteristic time $\tau_1$ if, in analogy to Eq.~(\ref{g3}), one considers
\begin{eqnarray}\label{periphery}
g_3(t) = \rho \langle \left(\frac{l}{N}\right)^2\rangle \sqrt{N}t,
\end{eqnarray}
in case that {\em only} the membrane periphery of length $\propto \sqrt{N}$
contributes to the center of mass displacement. Thus during $1 < t \le \tau_1$
the diffusion coefficient of the membrane becomes $D \propto N^{-3/2} \propto
L^{-3}$ reflecting the slow displacement of the center of mass.
Eq.~(\ref{periphery}) therefore suggests $\tau_1 \propto
L^{3+2\nu_{\text{min}}}\approx L^{4.4}$. With $g_1(\tau_1)\propto L^{\alpha
(3+2\nu_{\text{min}})}\approx \lambda_{\text{min}}^2$ one obtains then the
broken exponent for a {\em periphery-driven} membrane $\alpha_p =
(2\nu_{\text{min}})/(3+2\nu_{\text{min}}) = 0.32$.

In contrast, if only a finite number of loosely bound monomers at the vertices
effect the net displacement of the center of mass, one obtains then
\begin{eqnarray}\label{vertices}
g_3(t) = \rho \langle \left(\frac{l}{N}\right)^2\rangle t,
\end{eqnarray}
and therefore temporarily $D \propto N^{-2} \propto L^{-4}$ so that $\tau_1
\propto L^{4+2\nu_{\text{min}}}\approx L^{5.4}$. In this case one gets the
exponent for anomalous diffusion of a {\em vertices-driven} membrane
$\alpha_v = (2\nu_{\text{min}})/(4+2\nu_{\text{min}}) = 0.26$.

Thus we obtain two estimates which may be considered as the lower and upper
bounds of the anomalous diffusion exponent, $\alpha_v \le \alpha \le \alpha_p$,
depending
on the particular mechanism involved in the diffusive motion. The measured value
of $\alpha \approx 0.27$ lies indeed within these limits. Of course, one should
bear in mind that most probably neither mechanism of diffusion (vertices-driven,
or periphery-driven) takes place alone and the real process involves a mixture
of both. Moreover, at this point we cannot rule out the possibility that at
times $\tau_1 \le t \le \tau_R$ {\em all} membrane segments eventually get the
chance to perform an elementary  move and thus contribute to the center of mass
motion. Such a possibility would imply that during this time interval of
subdiffusive motion one observes a MSQD $g_1(t) \propto t^\alpha$ with an
exponent $\alpha = 0.5$, cf. Eq.~(\ref{beta}). We have indicated such a
behavior in Fig.~\ref{g1} by a dashed line with slope $0.5$ and it appears
compatible with the course of $g_1(t)$ in between $\tau_1$ and $\tau_R$ for our
largest membranes $L=30\div 50$. If such a diffusive regime really exists, it
would underline the multiscaling character of tethered membranes
\cite{Pandey,Pandey_old}. It is clear, however, that larger micelles
need to be simulated with satisfactory statistics before an unambiguous
conclusion in this respect can be drawn.

\section{Summary and conclusions}

In the present work we have studied the stochastic dynamics of flat
self-avoiding tethered membranes which are assumed to be completely permeable
to the surrounding good solvent and are thus expected to display typical Rouse
behavior. By means of extensive Monte Carlo simulations we find that the static
properties of our tethered membranes are described by scaling exponents which
agree very well with the appropriate theoretically predicted values. Thus the
radius of gyration scales with membrane linear size $L$ as $R_g^2 \propto
L^{2\nu}$ with $\nu = 1.02 \pm 0.01$, and the membrane thickness,
$\lambda_{\text{min}}^2 \propto L^{2\nu_{\text{min}}}$ with roughness exponent
$\nu_{\text{min}} = 0.70 \pm 0.01$ while the membrane asphericity vanishes
asymptotically as $A =\lambda_{\text{min}}^2 / \lambda_{\text{max}}^2 \propto
L^{-2\nu_A}$ with $\nu_A = 0.32 \pm 0.01$.

In the regime of Rouse diffusion we find with good accuracy that the diffusion
coefficient $D_N \propto N^{-1}$, as predicted, whereas the typical relaxation
time of such polymerized membranes grows as $\tau_R \propto L^4$ with the linear
dimension $L$. A comparative study, involving linear polymers too, reveals
also that the diffusion coefficient of permeable self-avoiding tethered
membranes, containing $N$ segments, is three times smaller than that of linear
polymer chains with the same number of segments.

Our main concern in this study, however, is with the subdiffusive motion of the
membranes central segment at times $t \le \tau_R$. Our numeric studies reveal
several regimes of anomalous diffusion whereby the central node MSQD grows as
$g_1\propto t^{0.86}$ for $t \le 1$, then $g_1 \propto t^{0.27}$ for $1 \le t
\le \tau_1$, further, with $g_1 \propto t^{0.5}$ at time $\tau_1
\le t \le \tau_R$, before turning eventually to normal diffusion with $g_1
\propto t$ for $t \ge \tau_R$. We use simple scaling arguments to interpret our
observation and suggest that the anomalous diffusion exponent $\alpha \approx
0.27$ which we find in agreement with recent studies \cite{Pandey,Pandey_old}
most probably reflects several particular mechanisms of membrane motion. These
mechanisms involve different groups of loosely bound membrane monomers whose
random hops predominantly contribute to the center of mass motion of the whole
membrane at times when most of the inner monomers are mutually blocked by their
nearest neighbors and, therefore, remain rather immobile. The particular
geometry of the membrane sheet (e.g., square, hexagonal, or rhombic) is expected
to enhance the role of either periphery, or vertex monomers, and therefore
slightly modify the observed value of the anomalous exponent $\alpha$ according
to Eqs.~(\ref{periphery}) and (\ref{vertices}). This would explain some small
deviations of our data from that of earlier
measurements~\cite{Pandey_old,Pandey}.

We believe that our results shed some light and provide insight into the complex
dynamics of polymerized membranes. It is, however, clear that further work is
needed before the nature of the membrane stochastic dynamics is definitely
established and understood.

\section{Acknowledgements}

The authors are indebted to the Max-Planck Institute for Polymer Research in
Mainz, Germany for hospitality during the stay of one of us (A.M.) as well as
for the possibility to use the computational facilities of the institute.


\begin{thebibliography}{99}
\bibitem{Gompper} G. Gompper and D. M. Kroll, J. Phys.: Condens. Matter {\bf
9}, 8795 (1997).
\bibitem{Schmidt} C. F. Schmidt, K. Svoboda, N. Lei, I. B. Petsche, L. E.
Berman, C. R. Safinya and G. S. Grest, Science {\bf 259}, 952 (1993).
\bibitem{Hwa} T. Hwa, E. Kokufuta, and T. Tanaka, Phys. Rev. A {\bf 44},
R2235 (1991).
\bibitem{Spector} M. S. Spector, E. Naranjo, S. Chiruvolu, and J. A.
Zasadzinski, Phys. Rev. Lett. {\bf 73}, 2867 (1994).
\bibitem{Sinsawat} A. Sinsawat, K. L. Anderson, R. A. Vaia, and B. L. Farmer, J.
Polym. Sci., Part B: Polym. Phys. {\bf 41}, 3272 (2003).
\bibitem{Vaia} {\em Polymer nanocomposites: Synthesis, Characterization, and
Modeling}, ed. R. Krishnamoorti and R. A. Vaia, ACS Symposium Series 804,
Washington D. C., 2002.
\bibitem{KKN} Y. Kantor, M. Kardar, and D. R. Nelson, Phys. Rev. A {\bf 35},
3056 (1987).
\bibitem{Doussal} P. Le Doussal, J. Phys. A: Math. Gen. {\bf 25}, L469 (1992).
\bibitem{Liu} D. Liu and M. Plischke, Phys. Rev. A {\bf 45}, 7139 (1992).
\bibitem{YKKK} Y. Kantor and K. Kremer, Phys. Rev. E {\bf 48}, 2490 (1993).
\bibitem{Radzihovsky} L. Radzihovsky and J. Toner, Phys. Rev. Lett. {\bf 75},
4752 (1995).
\bibitem{Plishke} M. Plischke and D. Boal, Phys. Rev. A {\bf 38}, 4943 (1988).
\bibitem{Boal} D. Boal, E. Levinson, D. Liu, and M. Plischke, Phys. Rev. A {\bf
40}, 3292 (1989).
\bibitem{FAbrah} F. F. Abraham, W. E. Rudge, and M. Plischke, Phys. Rev. Lett.
{\bf 62}, 1757 (1989).
\bibitem{DWH} C. M\"unkel and D. W. Heermann, Phys. Rev. Lett.
{\bf 75}, 1666 (1995).
\bibitem{Mori} S. Mori and S. Komura, J. Phys. A: Math. Gen. {\bf 29},
7439 (1996).
\bibitem{Drovetsky} B. Y. Drovetsky, J. C. Chu, and C. H. Mak, J. Chem. Phys.
{\bf 108}, 6554 (1998).
\bibitem{Kownacki} J.-Ph. Kownacki and H. T. Diep, Phys. Rev. E {\bf 66},
066105 (2002).
\bibitem{Koibuchi} H. Koibuchi, Z. Sasaki, and K. Shinohara, Phys. Rev. E {\bf
70}, 066144 (2004).
\bibitem{Grest} G. S. Grest and I. B. Petsche, Phys. Rev. E {\bf 50},
R1737 (1994).
\bibitem{GompperKroll} G. Gompper and D. M. Kroll, J. Phys.: Condens. Matter
{\bf 12}, A29 (2000).
\bibitem{Frey}E. Frey and D. R. Nelson, J. Phys. I France {\bf 1}, 1715 (1991).
\bibitem{Wiese} K. J. Wiese, Eur. Phys. J. B {\bf 1}, 269 (1998).
\bibitem{Pandey} R. B. Pandey, K. L. Anderson, and B. L. Farmer, Phys. Rev. E
{\bf 75}, 061913 (2007).
\bibitem{Pandey_old} R. B. Pandey, K. L. Anderson, H. Heinz, and B. L. Farmer,
J. Polym. Sci., Part B: Polym. Phys. {\bf 43}, 1041 (2005); {\bf 43}, 3478
(2005); {\bf 44}, 2512 (2006).
\bibitem{Bird} R. B. Bird, C. F. Curtiss, R. C. Armstrong, and O. Hassager,
Dynamics of Polymeric Liquids, 2nd ed., Vol. 2 (Wiley, New York, 1987).
\bibitem{JCP} H. Popova and A. Milchev, J. Chem. Phys. {\bf 127}, 194903 (2007).
\bibitem{AMKBJB} K. Binder, A. Milchev, and J. Baschnagel, Annu. Rev. Mater.
Sci. {\bf 26}, 107 (1996).
\end{thebibliography}
\end{document}